\documentclass[aps,prl,superscriptaddress,showpacs,twocolumn]{revtex4-1}

\usepackage{graphicx}
\usepackage{float}
\usepackage{amsmath,amssymb}
\usepackage{color}
\usepackage[colorlinks,citecolor=blue,linkcolor=blue,urlcolor=blue]{hyperref}

\begin{document}
\preprint{March 2019}

\title{Single-atom control of the optoelectronic response in sub-nanometric cavities}
\author{P. Garc\'ia-Gonz\'alez}
\affiliation{Departamento de F\'{\i}sica Te\'orica de la Materia Condensada and Condensed
Matter Physics Center (IFIMAC), Universidad Aut\'onoma de Madrid, E-28049
Cantoblanco, Madrid, Spain}
\email{pablo.garciagonzalez@uam.es}
\affiliation{Nano-Bio Spectroscopy Group and ETSF, Universidad del Pa\'{\i}s Vasco UPV/EHU, Avenida de Tolosa 72, E-20018 Donostia, Spain}
\author{Alejandro Varas}
\affiliation{Nano-Bio Spectroscopy Group and ETSF, Universidad del Pa\'{\i}s Vasco UPV/EHU, Avenida de Tolosa 72, E-20018 Donostia, Spain}
\author{F. J. Garc\'ia-Vidal}
\affiliation{Departamento de F\'{\i}sica Te\'orica de la Materia Condensada and Condensed
Matter Physics Center (IFIMAC), Universidad Aut\'onoma de Madrid, E-28049
Cantoblanco, Madrid, Spain}
\affiliation{Donostia International Physics Center (DIPC), Paseo Manuel Lardizabal 4, E-20018 Donostia, Spain}
\author{Angel Rubio}
\affiliation{Max Planck Institute for the Structure and Dynamics of Matter, Center for
Free-Electron Laser Science, and Department of Physics, Luruper Chausse 149,
22761 Hamburg, Gemany}
\affiliation{Nano-Bio Spectroscopy Group and ETSF, Universidad del Pa\'{\i}s Vasco UPV/EHU, Avenida de Tolosa 72, E-20018 Donostia, Spain}
\affiliation{Donostia International Physics Center (DIPC), Paseo Manuel Lardizabal 4, E-20018 Donostia, Spain}
\affiliation{Center for Computational Quantum Physics, Flatiron Institute, 162 5th Ave. New York, NY 10010}
\date{\today}

\begin{abstract}
By means of ab-initio time dependent density functional theory calculations carried out on an prototypical 
hybrid plasmonic device (two metallic nanoparticles bridged by a one-atom junction), we demonstrate the strong 
interplay between photoinduced excitation of localized surface plasmons and electron transport through the 
single atom. Such an interplay is remarkably sensitive to the atomic orbitals of the junction. 
Therefore, we show the possibility of a twofold tuning (plasmonic response and photoinduced current across the juntion)
just by changing a single atom in the device.
\end{abstract}

\maketitle

The interface between Plasmonics and electron transport phenomena is becoming an intense area 
of research~\cite{Ward10,Arielly11,Savage12,Scholl13,Hajisalem14,Vadai13,Tan14,Benz15,Lerch16,Sanders16,Xiang17,Lerch17}. 
This is mainly motivated by the ability of localized surface plasmons to concentrate light in sub-nanometric
``hot spots'' in a controllable 
manner~\cite{Benz16b,Barbry15,Urbieta18,Shin18}, which offers 
unique opportunities in the design of novel molecular-scale optoelectronic devices
exhibiting highly tunable operativeness.
Systems comprising plasmonic nanostructures bridged by tunnel junctions
constitute the natural route towards the realization of such devices.
Among those, 
simple \textit{vacuum} nanogaps in metallic nanoparticle dimers have been addressed 
experimentally~\cite{Savage12,Scholl13,Hajisalem14}, and analyzed theoretically using 
methods that account for the quantum nature of the 
electron dynamics~\cite{Zuloaga09,Marinica12,Esteban12}, and also for 
the specific atomic structure of the system~\cite{Zhang14,Barbry15,Varas15,Varas16}. 
Then, the interplay between the plasmonic response and the
across-gap photoinduced current is presently well understood.
More interesting is the case of hybrid systems where metallic
nanostructures are bridged by atomic or molecular 
junctions~\cite{Ward10,Arielly11,Vadai13,Tan14,Benz15,Lerch16,Sanders16,Xiang17,Lerch17},
since the electron orbitals of the junction also contributes to the current~\cite{Galperin12}. 
In fact, it has been proposed that even a single-atom junction
located in a plasmonic cavity can modify dramatically its optical response~\cite{Emboras16}.
Consequently, mechanically-induced discontinuous changes on the composition
of multiple-atom junctions will be reflected accordingly on both the optical
absorption spectrum and the conductance of the junction~\cite{Rossi15,Marchesin16}.

\begin{figure}[t!]
\includegraphics[width=0.95\linewidth]{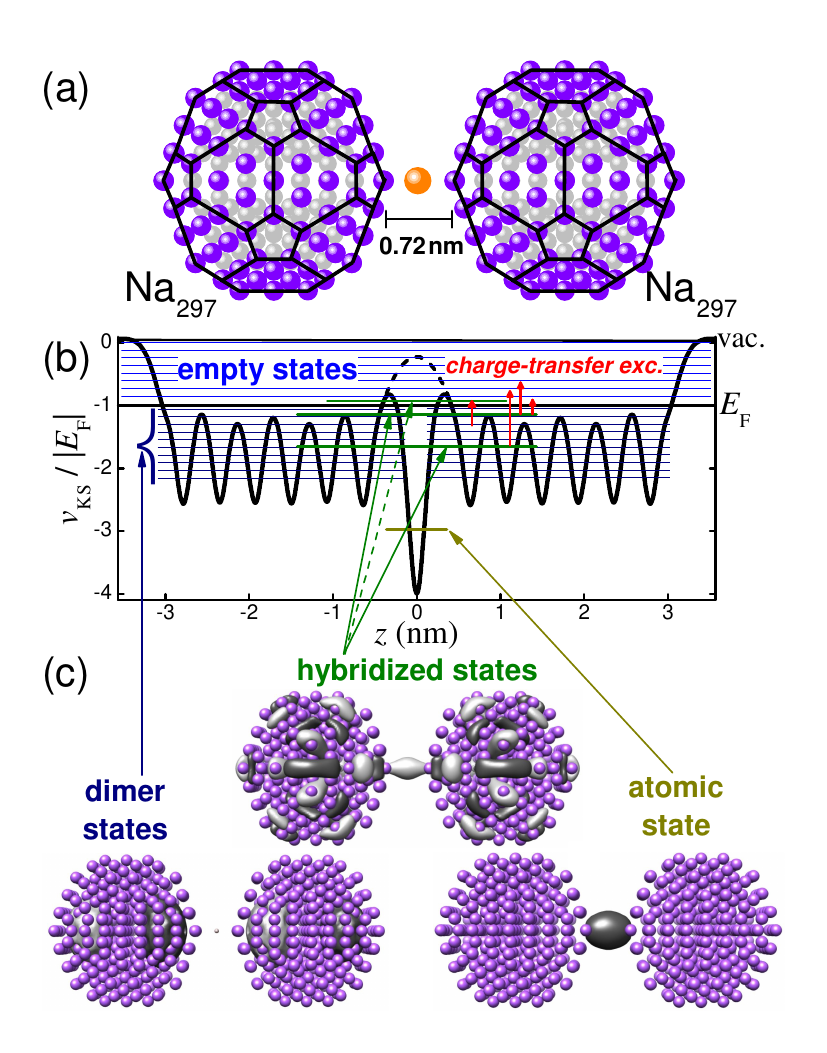}
\caption{(a) Representation of the geometrical arrangement of an
icosahedral-Na$_{297}$ dimer bridged by a single atom. (b) Schematic
representation of the KS one-electron energy leveles and the KS potential
along the dimer axis (solid line) compared with the KS potential for the
bare dimer (dashed line). (c) Isosurface plot of the three types of obitals
appearing in the $\mathrm{Na}_{297}-\mathrm{Al}-\mathrm{Na}_{297}$ system:
dimer, bound atomic, and hybridized states.}
\label{fig:1}
\end{figure}

Theoretical analyses of these hybrid plasmonic 
devices~\cite{Rossi15,Marchesin16,Song11,Song12,Kulkarni15} are by far not as abundant 
as for their vacuum nanogaps counterparts, and
a full understanding of the underlying excitation mechanisms is still lacking. 
In this Letter we provide simple but robust explanations of the most relevant
phenomena, which are assessed by means of ab-initio 
calculations of a prototypical hybrid plasmonic system: 
two metallic $\mathrm{Na}_{297}$ clusters bridged by a single atom (see Fig.~\ref{fig:1}). 
We show that, due to the high sensitivity of the photoinduced current 
to the position of the energy levels of the junction, 
the optoelectronic response of the nanodevice can be indeed tuned 
by such a single atom.
 
The two clusters $\mathrm{Na}_{297}$ are in their stable icosahedral arrangement~\cite{Noya07}
and orientated facing three-atom edges separated by a distance $d=0.72\ $nm.
Therefore, the system is symmetric with respect to the $XY$ plane, $OZ$
being the dimer axis and $OX$ the axis parallel to the facing edges. 
For this separation the overlap of the
ground-state densities of the two clusters is negligible 
[for the ``bare'' $\mathrm{Na}_{297}$ dimer, 
the minimum of the potential barrier on the $XY$ plane is 
$\Delta _{\mathrm{b}}=1.35$ eV above the Fermi level 
($E_{\mathrm{F}}=-2.85\ \mathrm{eV}$)].

Well-converged Kohn-Sham (KS) calculations~\cite{Kohn65} using standard 
norm-conserving pseudopotentials~\cite{Troullier91} and the spin-dependent generalized gradient 
approximation~\cite{Perdew96}, are carried out on a real space representation 
(grid spacing of 0.026 nm) using the OCTOPUS code~\cite{Marques03,Castro06,Andrade15}. 
Then, the linear optical response of the system to an incident monochromatic
electromagnetic field polarized along the dimer axis is obtained in the quasi-static 
approximation using time dependent density functional  theory (TDDFT)~\cite{Runge84}. 
Namely, since the
wavelength of the incident light is much larger than the size of the
nanosystem, the perturbing electric field is $\mathbf{E}_{\omega }(\mathbf{r},t)
\simeq E_{0}\exp (-\mathrm{i}\omega t)\mathbf{e}_{z}$. The 
frequency-dependent response is evaluated from a single propagation of the TDDFT
Runge-Gross equations considering that the system is perturbed at $t=0$ by a
delta-kick electric field $\mathbf{E}_{\mathrm{D}}(\boldsymbol{r},t)=
E_{0}\tau _{0}\delta (t)\mathbf{e}_{z}$~\cite{Yabana96}. From the corresponding induced
electron density $\delta n(\mathbf{r},t)$ we can calculate the$\ z$%
-component of the induced electric dipole, $\delta \mathcal{D}_{z}(t)$,
and the induced charge in the $z>0$ region of the system, $\delta Q_{+}(t)$.
Then, the absorption cross section is
given by $\sigma _{\mathrm{abs}}(\omega )=-\omega \alpha(\omega
)/(4\pi c)$, where 
\begin{equation}
\alpha (\omega )=\frac{1}{E_{0}\tau _{0}}\int_{0}^{+\infty }dt\ \delta 
\mathcal{D}_{z}(t)e^{+\mathrm{i(}\omega +\mathrm{i}\gamma )t}  \label{Eq:01}
\end{equation}
is the frequency-dependent polarizability, and $\gamma $ is a damping frequency accounting for non-electronic 
dissipation mechanisms. On the other hand,
\begin{equation}
I(\omega )=\frac{1}{\tau _{0}}\int_{0}^{+\infty }dt\ \left[ \frac{d}{dt}%
\delta Q_{+}(t)\right] e^{+\mathrm{i(}\omega +\mathrm{i}\gamma )t}\
\label{Eq:02}
\end{equation}
is the intensity of the induced AC current across the junction.
In practice, the time propagation is truncated at a time $T_{\max }$, and well-resolved results
are achieved by using $\gamma \simeq 35\ \mathrm{meV}$, $T_{\max }\simeq 40\ 
\mathrm{fs}$, and a time step $\Delta t\simeq 0.002\ \mathrm{fs}$.

\begin{figure}[t!]
\includegraphics[width=0.95\linewidth]{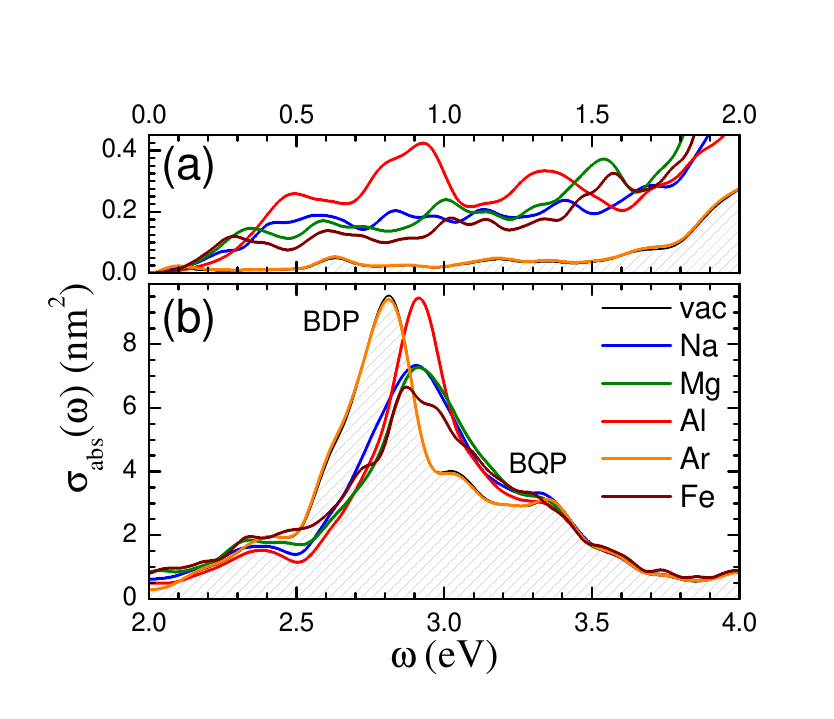}
\caption{Dependence of the optical response on the atomic element
that constitutes the junction for bridged $\mathrm{Na}_{297}$ dimers.
(a) Optical absorption cross section in the low frequency range ($0 - 2$
eV). (b) As in panel (a) in the ``plasmonic'' frequency range $2 - 4$ eV. BDP and BQP stand for the 
bonding dipolar and quadrupolar surface plasmons, respectively (see text). Note the different scale in panel (a).}
\label{fig:2}
\end{figure}

\begin{figure}[t!]
\includegraphics[width=0.95\linewidth]{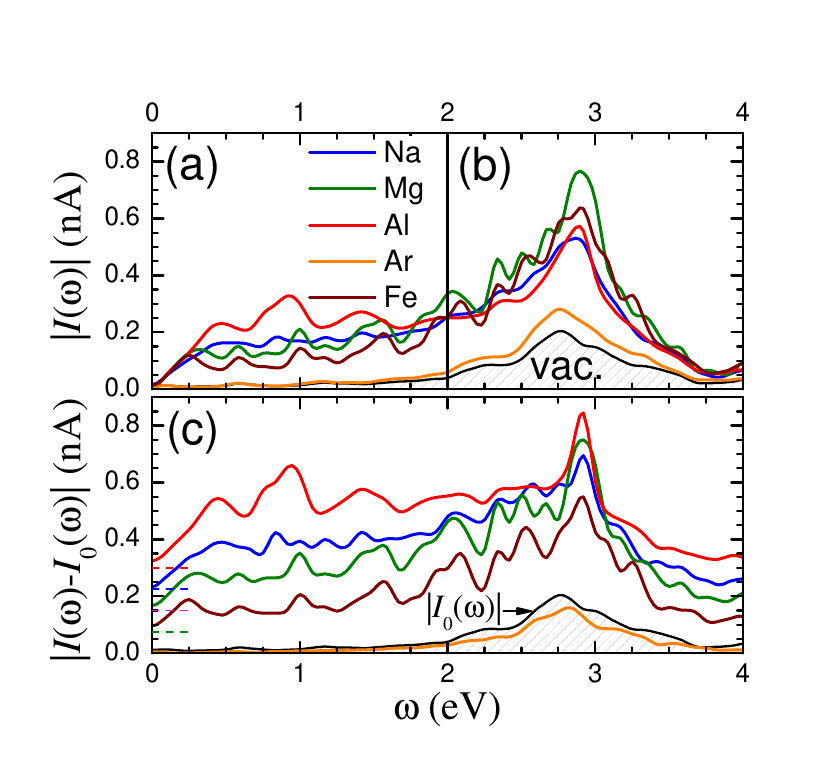}
\caption{Frequency dependence of the photoinduced current intensity across the one-atom junction 
by an incident E-field ($E_0=10^6~\mathrm{V/m})$. (a,b) Amplitude $| I (\omega) |$. 
(c) Waterfall representation of the junction contribution to the current (see text), 
$| I (\omega) -  I_0 (\omega)|$.}
\label{fig:3}
\end{figure}

The calculated optoelectronic response is depicted in Fig.~\ref{fig:2} and \ref{fig:3} 
for different one-atom junctions (Na, Mg, Al, Ar, and Fe) and compared to the
response of the bare dimer.
In line with previous jellium~\cite{Zuloaga09,Marinica12} and ab-initio~\cite{Zhang14,Varas15} 
TDDFT calculations,
the optical absorption for the $\mathrm{Na}_{297}$ dimer (Fig.~\ref{fig:2}) is dominated by the 
so-called bonding dipolar plasmon (BDP) at $\omega _{\mathrm{p}}=2.8\ \mathrm{eV}$, 
whereas a second mode, the bonding quadrupolar plasmon (BQD), appears as a less-defined 
spectral feature at $\omega \sim 3.3\ \mathrm{eV}$. 
These plasmonic features emerge after a high renormalization of low-energy single-electron 
excitations due to the electron-electron (e-e) interaction~\cite{Yannouleas93,Bernadotte13,Zhang17}. 
The induced AC current across the empty cavity (Fig.~\ref{fig:3}) is very weak 
in the low-frequency region, since it is established by excitations 
between bonding and antibonding dimer states below the potential barrier. 
However, for energies greater than $\Delta _{\mathrm{b}}$ 
the current is mainly established through excitations to unoccupied states above the 
potential barrier (including virtual resonant states above the
vacuum level), which are strongly enhanced by the induced $E$-field associated to a
plasmon resonance.

The above picture changes dramatically for single-atom junctions.  
As it is schematically depicted in Fig. \ref{fig:1}, the presence of the atom leads to 
a depletion of the potential barrier, as well as to the 
appearance of deeply-bound atomic states and occupied and unoccupied hybridized 
atom-dimer states. Single-electron excitations involving one of these new hybridized 
orbitals are much less renormalized by the e-e interaction than the 
ones between dimer states. As a consequence, under an independent electron picture
(i.e., neglecting e-e interactions in the excitation process) those transitions
can hardly be distinguished in the absorption spectrum. However, once 
e-e interactions are considered, the series of low-energy excitations from an 
occupied hybridized state to several unoccupied dimer states (or from several occupied dimer states
to an unoccupied hybridized state) are not obscured anymore by nearby excitations between dimer states. 
Then, they become visible (not embedded in the quasi-continuum of dimer excitations) as 
identifiable features in the spectrum at frequencies
$\omega \simeq \omega_0 + \langle \phi_{\mathrm{HS}}\phi_{\mathrm{dim}} | 
\hat{w}_\mathrm{c}+\hat{K}_{\mathrm{xc}} | \phi_{\mathrm{dim}}\phi_{\mathrm{HS}}\rangle$
having an oscillator strength 
$f \simeq \omega | \langle  \phi_{\mathrm{dim}} | \hat z | \phi_{\mathrm{HS}}\rangle |^2$,
$\omega_0$ being the KS excitation energy 
(i.e., the independent-electron transition energy), $\hat{w}_\mathrm{c}$ 
the Coulomb interaction operator, and $\hat{K}_{\mathrm{xc}}$ the so-called XC kernel~\cite{Appel03}. 
That is, although these modes are not plasmonic at all, they appear in the spectrum thanks to the
plasmonic response of the system.
Bearing in mind the quasi-continuous DOS of the dimer states and that 
one atomic orbital of energy $\epsilon_{\mathrm{at}}$ leads to 
several hybridized states around $\epsilon_{\mathrm{at}}$, once the lifetime of the excitations is 
accounted for, the expected outcome from each atomic orbital is a series of asymmetric 
Fano-like peaks. The overall structure of each series will depend on the DOS of occupied [unoccupied] 
dimer states if the hybridized state that is involved in the transitions is unoccupied [occupied].

These excitations are naturally associated to a net charge transfer (CT) between the clusters. 
For the case of an incident low-frequency E-field $\mathbf{E}_{\omega }(\mathbf{r},t)$ 
in resonance with a transition between the states $| \phi_{\mathrm{HS}} \rangle$ 
and $| \phi_{\mathrm{dim}} \rangle$, the driven current across the junction is in phase with the
external field, whereas its intensity is
$I(t) \propto E_0\exp(-\mathrm{i}\omega t)
\langle \phi_{\mathrm{HS}} | \hat{z} | \phi_{\mathrm{dim}} \rangle
\int_{z >0} \phi_{\mathrm{HS}}(\mathbf{r})\phi_{\mathrm{dim}}(\mathbf{r}) d^3\mathbf{r}$.
As a consequence, the amplitude of the photoinduced density, $|I(\omega)|$,
will follow an oscillatory behavior in the low-frequency regime resembling the one corresponding to
the optical absorption. These are indeed the trends shown by the ab-initio calculations as can be seen 
in panels (a) of Fig.~\ref{fig:2} and \ref{fig:3}. Then, it is not a surprise that the changes
induced by an Ar atom are negligible, whereas the optoelectronic response for the rest of the cases 
depends very sensitively on the atom that constitutes the junction.

As mentioned above, this low-frequency response is mainly determined by the DOS of the system. 
The eigenenergies of the hybridized states can be easily extracted by direct comparison of the DOSs
of the bridged- and the bare-dimer systems. For the Mg and Fe cases, the occupied s-like
hybridized states lie well below the Fermi level (around 2.3 and 2.7 eV below $E_\mathrm{F}$ for 
Mg and Fe, respectively; the Fe $3\mathrm{d}^6$ atomic states preserving their atomic character), 
whereas the unoccupied hybridized states appear for both cases around 0.3 eV above $E_\mathrm{F}$.
Therefore, the CT modes correspond to transitions from occupied dimer states to
unoccupied hybridized states. Since the hybridization is larger for the embedded Mg atom, the 
absorption is more intense in this case, but their overall spectral shape is very similar. 
For the Na junction, there are occupied and unoccupied 3s hybridized states around the Fermi level. Then, the optical
absorption spectrum results from the superposition of two series of peaks, corresponding to transitions 
from [to] occupied [unoccupied] hybridized states. Finally, for Al we have several channels for the CT modes: 
occupied and unoccupied $3\mathrm{p}_z$ hybridized states close to $E_\mathrm{F}$, 
as well as unoccupied $3\mathrm{p}_x$ ones. Due to the number of channels
and the efficient hybridization of the p orbitals, the optical absorption for the Al junction is more intense
but less resolved.

CT modes do also contribute to the higher-frequency response, which is primarily
dominated by the BDP of the dimer. Since CT excitations involve a dimer state, 
they will be necessarily hybridized with the renormalized single-electron excitations that constitutes the surface plasmon. 
For the systems considered in this Letter, a detailed quantitative analysis of such 
hybridizations is cumbersome. However, if atomic-chain dimers are considered instead, 
the plasmon is made up by only one (or a few) highly renormalized single-electron 
excitations~\cite{Bernadotte13}. In this case, the strong hybridization of 
CT and plasmon modes is evident~\cite{Barquilla19} and the final plasmonic feature is made up by 
two CT excitations with a mixed \textquotedblleft plexcitonic\textquotedblright character~\cite{Manjavacas11}.
Moreover, new CT channels involving deeper occupied hybridized states will now contribute to the response. 
The AC current across the junction will be then enhanced by the participation 
of hybridized CT modes in the plasmonic response and, to a lesser extent, by the aforementioned 
depletion of the potential barrier between the clusters. 
As a consequence, a redistribution of the spectral weight around the original BDP 
frequency is expected, which results on changes of the width and shape of the surface
plasmon peak. In addition, due to the enhanced current between the nanoparticles, 
the BDP will be blueshifted~\cite{PerezGonzalez10}. 

\begin{table}[t!]
\centering
\begin{tabular}{l c c c c c c c c c c c}
\hline
\hline
Atom & vac. & Na & Mg & Al & Si & P & S & Cl & Ar & Fe \\
\hline
$\omega_\mathrm{p}$  (eV) & 2.81 & 2.91 & 2.91 & 2.92 & 2.89 & 2.88 & 2.87 & 2.85 & 2.81 & 2.88 \\
$\Delta\omega_\mathrm{p}$  (eV)   & 0.31 & 0.55 & 0.49 & 0.32 & 0.39 & 0.37 & 0.32 & 0.31 & 0.31 & 0.58 \\
\hline
\hline
\end{tabular}
\caption{Frequency ($\omega_\mathrm{p}$) and width ($\Delta\omega_\mathrm{p}$) 
of the main plasmon resonance for a $\mathrm{Na}_{297}$ dimer bridged by a single atom. Note the sensitivity of the width on the atomic element that constitutes the junction.}
\label{tab:1}
\end{table}

The above considerations perfectly explain the results  depicted
in panels (b) of Fig.~\ref{fig:2} and \ref{fig:3} [see also Table~\ref{tab:1}, 
where we present the maximum and the width (FWHM) of the plasmon 
peak for different atomic junctions].
The inclusion of an Ar atom does not change at all the optical absorption, 
but the intensity of the induced current increases for $\omega > 1$ eV range 
due to the depletion of the potential barrier. However, for the rest of
the junctions there is a noticeable modification of the plasmonic response 
which depends on the specific atomic element. In general, the main BDP 
feature is blueshifted, broadened, and for the Fe case also fragmented. 

This dependence plenty manifests in the behavior of the induced current.
A fair estimation of the changes in the current due to the 
one-atom junction itself is given by $| I (\omega) - I_0 (\omega)|$, 
where $I_0 (\omega)$ is the intensity of the induced current for the bare 
$\mathrm{Na}_{297}$ dimer. As we may see in panel (c) of Fig.~\ref{fig:3},
there is a plasmon-induced strong enhancement of the CT-modes contribution 
to the current. Besides, for the Mg- and Fe-bridged systems there are new oscillatory 
features which correspond to higher-frequency series of CT 
excitations from occupied s-like hybridized states of the Mg and Fe atoms. 
In fact, for the case of Fe the hybridization of
one of these CT excitations is so intense that it can be distinguished not only in the AC current spectrum 
but also in the optical absorption, being the main responsible of the aforementioned split of the BDP. 

\begin{figure}[t!]
\includegraphics[width=0.95\linewidth]{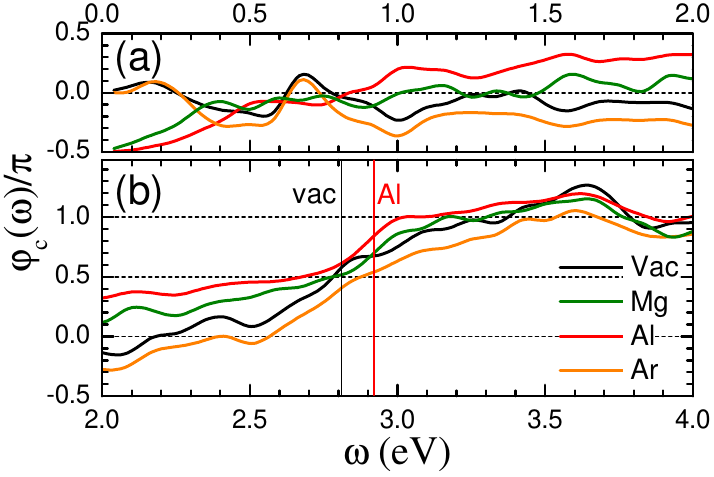}
\caption{Unwrapped phase shift $\varphi_\mathrm{c}(\omega)$
of the photoinduced AC current across the one-atom junction: 
(a) low frequency range ($0 - 2$ eV); (b) ``plasmonic'' frequency range ($2 - 4$ eV).
The black and red vertical lines in Panel (b) are located at the BDP plasmon frequency of
the clean and the Al-bridged dimer, respectively. The cualitative behaviors of $\varphi_\mathrm{c}(\omega)$
for Na- and Fe-bridged dimer are very similar than the ones corresponding to Al- and Mg-bridged dimers,
respectively.}
\label{fig:4}
\end{figure}

Finally, the existence of CT modes is also reflected on the
AC-current phase $\varphi_\mathrm{c}(\omega) = \arctan [\Im I(\omega) / \Re I(\omega)]$ (see Fig.~\ref{fig:4}).
When such modes do not exist (as in the bare and Ar-bridged dimers), 
the induced current in the static limit ($\omega \rightarrow 0$) is in phase with the external E-field, which is 
the expected behavior of a ``pure'' tunneling regime. Then, in the low-frequency range, where the driving E-field 
is mainly the external one, $\varphi_\mathrm{c}(\omega)$ oscillates around zero, although the center of the oscillations
is shifted to negative values for higher frequencies. Since this shift is more pronunced for the Ar-bridged
dimer and appears at lower frequencies than for the bare dimer, we may attribute it to the opening of new excitation
channels above the potential barrier betwen the two clusters. Finally, in the ``plasmonic'' frequency range,
where the driving E-field is mainly the induced one, the phase increases almost monotonically from 
$\varphi_\mathrm{c}(\omega) \simeq 0$, in such a way that $\varphi_\mathrm{c}(\omega) \simeq \pi/2$ 
at the BDP frequency~\cite{Marinica12,Varas16} and $\varphi_\mathrm{c}(\omega) \simeq \pi$ at the BQP frequency.

By contrast, when the current is mainly established by excitations of CT modes (for instance, Mg- and Al-bridged dimers),
$\varphi_\mathrm{c}(\omega) \simeq -\pi/2$ in the static limit, as correponds to an out-of-resonance transition. 
Then, the phase increases monotonically to a value approximately equal
to zero until the onset of CT excitations is reached and, eventually, it slowly increases in an oscillatory manner to a 
value $\varphi_\mathrm{c}(\omega) \simeq \pi/2$ around the \textit{bare-dimer} BDP plasmon frequency 
$\omega_\mathrm{c}^\mathrm{(b)}$. Finally, the behavior for $\omega > \omega_\mathrm{c}^\mathrm{(b)}$
as the frequency increases is qualiatively simlar than in the previous case. However, it is worth emphasizing that the phase 
$\varphi_\mathrm{c}(\omega)$ at the corresponding BDP frequency is closer to $\pi$ than to $\pi/2$, which is the most
noticeable outcome of the existence of CT excitations in this frequency range.

In conclusion, we have elucidated and illustrated by means of ab-initio calculations 
the mechanisms of the optoelectronic response in hybrid systems composed by metallic 
plasmonic nanoparticles bridged by atomic junctions.
We have shown how the changes in the optical and transport properties of the system 
depend strongly on the nature of the single atom junction but, unlike the DC regime in the limit of zero bias, 
the induced current is also affected by the overall DOS of the nanoparticles. Furthermore, the 
AC current can be enhanced through the hybridization of surface plasmon resonances
and charge-transfer modes.
Thus, the control of various parameters, such as the geometry
of the nanoparticles, their separation, type of molecules constituting the junction,
and also the dielectric environment, would offer us unique ways towards new 
molecular optoelectronic devices.

\begin{acknowledgments}
Enlightening discussions with J.~C. Cuevas, A.~I. Fern\'andez-Dom\'{\i}nguez, and C. Tejedor 
are thankfully acknowledged. 
This work has been supported by the European Research Council 
(ERC-2011-AdG-290981 and ERC-2015-AdG-694097)
and the Spanish Government 
(``Mar\'{\i}a de Maeztu'' Programme for Units of Excellence in R\&D MDM-2014-0377 
and grant MAT2014-53432-C5-5-R).
\end{acknowledgments}

\end{document}